\documentclass[conference]{IEEEtran}

\usepackage[english]{babel}
\usepackage[utf8x]{inputenc}
\usepackage{amsmath}
\usepackage{graphicx}
\usepackage[colorinlistoftodos]{todonotes}
\usepackage{amsthm}
\usepackage{algorithm}
\usepackage[noend]{algpseudocode}
\usepackage{subfigure} 
\usepackage[normalem]{ulem}
\usepackage{multirow}
\usepackage{amssymb}

\addto\captionsenglish{}

\title{Reducing BESS Capacity for Accommodating Renewables in Subtransmission Systems with Power Flow Routers}
\author{\IEEEauthorblockN{Tianlun Chen,  Albert Y.S. Lam, Yue Song, and David J. Hill}
\IEEEauthorblockA{Department of Electrical and Electronic Engineering\\
The University of Hong Kong\\
Pokfulam Road, Hong Kong\\
Email: \{tlchen, ayslam, yuesong, dhill\}@eee.hku.hk}}

\begin{document}
\maketitle

\begin{abstract}
Widespread utilization of renewable energy sources (RESs) in subtransmission systems causes serious problems on power quality, such as voltage violations, leading to significant curtailment of renewables. This is due to the inherent variability of renewables and the high R/X ratio of the subtransmission system. To achieve full utilization of renewables, battery energy storage systems (BESSs) are commonly used to mitigate the negative effects of massive fluctuations of RESs. Power flow router (PFR), which can be regarded as a general type of network-side controller, has also been verified to enhance the grid flexibility for accommodating renewables. 
In this paper, we investigate the value of PFR in helping BESSs for renewable power accommodation. The performance of PFR is evaluated with the minimum BESS capacity required for zero renewable power curtailment with and without PFRs.
The operational constraints of BESSs and the terminal voltage property of PFRs are considered in a multi-period optimization model. 
The proposed model is tested through numerical simulations on a modified IEEE 30-bus subtransmission system and a remarkable result shows that 15$\%$ reduction of BESS capacity can be achieved by installing PFRs on a single line.
\end{abstract}


%\vspace{-0.2cm}
\section{Introduction} \label{sec:introduction}
\IEEEPARstart{R}{enewable} energy sources (RESs) bring many challenges to modern power system operators. For example, distributed energy resources (DERs), such as wind farms and solar photovoltaics (PVs), have a great impact on power quality, especially over-voltage and under-voltage problems in the subtransmission system. It is due to the high R/X ratio, which is the inherent property of subtransmission system and it makes the voltages more sensitive to active power permutation \cite{sun2019review}.
%Other problems like network flow congestion and temporal mismatch of power balance also restrict the penetration level of RESs \cite{hozouri2014use}. 
To reduce the negative effects of renewable power spillage while fulfilling the operational constraints, various control strategies and control devices have been developed to address the problems such as transmission network expansion plans (TNEP), on-load tap changers (OLTCs), and so forth \cite{sun2019review,procopiou2014voltage,deshmukh2012voltage}. The grid reinforcement can effectively increase the wind power injections but the cost-performance ratio is rather low and the required time is too long \cite{hozouri2014use}. In view of a multi-period time horizon, OLTCs may not be sufficient to handle the over-voltage and under-voltage problems at different buses simultaneously and timely. Based on the IEEE Standard 1547-2018 \cite{photovoltaics2018ieee}, both active and reactive powers from distribution generators (DGs) can be coordinated to maintain the voltages within an acceptable range. Consider a simple 2-bus system, the required reactive power support $\Delta Q$ is much more than the active power injection $\Delta P$ based on the DistFlow equation $\Delta Q = -\Delta P \cdot R/X $, where $R$ and $X$ are the resistance and reactance of the line \cite{baran1989network}. Renewable power curtailment or load shedding may occur to prevent voltage violations if the reactive power support is not sufficient. Moreover, the system operator may not be able to directly control all the DG inverters.

Recently, an alternative way to achieve full utilization of renewables is to use battery energy storage systems (BESSs). BESSs have been verified to effectively reduce the renewable power spillage and mitigate volatge variations \cite{dui2017two}. In \cite{dui2017two}, a two-stage optimization method is proposed to optimize the BESS capacity and power associated with the power dispatch schedule with the consideration of grid-connected wind farms. In \cite{zheng2013optimal}, the optimal allocation of BESSs for mitigating the risk of the distribution utility operation is investigated.  A distributed control architecture of BESSs for voltage regulation is proposed in \cite{zeraati2016distributed} for replacing the renewable power curtailment.  Moreover, a MPC-based program is applied for BESS operation to keep voltages within the acceptable range with high penetration of DERs \cite{zarrilli2017energy}.

However, the cost of installing BESSs in the system is still high. Power electronic-based devices have grown dramatically in the recent decades, which brings many opportunities to modern power grid development.
Power flow router (PFR) is proposed as a grid-level architecture for managing branch power flows \cite{lin2014architectural,chen2018transient}. The load flow model of PFR is developed to illustrate its capability on enlarging the feasible region of OPF problems. In \cite{lin2017optimal}, the generic architecture of PFR is proposed and its capability on available transfer capability (ATC) enhancement is verified. 
%In \cite{chen2018transient}, PFRs were applied to an optimal power flow problem with transient stability constraints, where the system stability is improved by installing PFRs on suitable lines. 
%From the perspective of terminal characteristic for voltage regulation, PFR can also be regarded as a typical FACTs device, which can avoid power flow congestion and accommodate the variations of RESs by its fast response feature.  

In this paper, we investigate the value of PFRs on replacing BESSs for accommodating renewables. We propose a novel optimal power flow (OPF)-based model for minimizing the BESS capacity with the help of PFRs while ensuring zero renewable curtailment. A predefined time horizon, e.g., 24 hours, is applied to the multi-period optimization problem, where the charging/discharging operations of BESSs satisfy the capacity limits and the required specifications of voltages, branch power flows, and PFRs. PFR is modelled as a series voltage regulator, which can control the terminal voltage magnitude and angle. The subtransmission system is described by a full AC power flow model for accuracy concerns, which induces non-convexity. A modified semi-definite programming (SDP)-based relaxation based on the bus injection model is adopted to convexify the non-convex problem with global optimum guarantee. The proposed methods are verified on a modified IEEE 30-bus system and a significant decrease of BESS capacity required for accommodating renewables can be achieved by installing a small number of PFRs.

The remaining parts of this paper are organized as follows.
In Section \ref{sec:problem formulation}, the system components, including PFRs and BESSs, and the OPF-based formulation are presented. Section \ref{sec:methods} illustrates the principles of convex relaxation and the modified SDP relaxation method to reformulate the original model. In Section \ref{sec:case study}, a modifed IEEE 30-bus subtransmission system is described and the proposed model is validated by a case study. Finally, we conclude this paper in Section \ref{sec:conclusion}. 
 
\section{Problem Formulation} \label{sec:problem formulation}
Consider a subtransmission system with the set of buses $\mathcal{N}= \{1,2,...,N \}$ and the set of lines $\mathcal{L}= \{(i,j) \} \subset \mathcal{N} \times \mathcal{N} $. Bus $1$ is assumed as the slack bus and the interconnection with the transmission network. The set of neighbors of each bus $i \in \mathcal{N}$ is denoted as $\Omega_i \subset \mathcal{N}$. 
For each bus $i \in \mathcal{N}$, $V_i^t \in \mathbb{C}$ is the complex voltage at time $t$, and $P_i^t$ and $Q_i^t$ are the active and reactive power injections at time $t$, respectively. 
%For each line $(i,j) \in \mathcal{L}$, the line admittance is represented by $y_{ij} = 1/z_{ij} =(r_{ij} + jx_{ij})^{-1}$, where $z_{ij}$, $r_{ij}$ and $x_{ij}$ denote the line impedance, resistance and reactance, respectively. Let $c_{ij} = c_{ji}$ be the shunt admittance along line $(i,j)$.
Let $Y$ denote the power network admittance matrix which includes the series and shunt components of lines. Denote $Y_{ij}$ as the $(i,j)$ entry of $Y$.
Denote $\mathcal{N}_W$ as the buses installed with renewable farms, $\mathcal{N}_G$ as the set of synchronous generators and $L$ as the total number of lines. 
%and $\mathcal{N}_E$ as the set of buses with BESSs.

Consider the optimization model in a time-slot basis and the time horizon is defined as the time slots $\mathcal{T} = \{1, \ldots, T\}$ with the time interval $\Delta T$. In this work, the time horizon $T$ is set as 24 hours and the time interval is 1 hour.
For $i \in \mathcal{N} $, $E_i^t$ represents the amount of stored energy contained in the BESS at bus $i$ in time slot $t$. The dynamics of BESS at bus $i$ can be modelled by the following first-order difference equation:
\begin{align}
    E_i^{t+1} =  E_i^{t} + P_{BESS,i}^{t} \Delta T, t\in \mathcal{T},    \label{SOC-1}
\end{align}
where $P_{BESS,i}^t$ represents the charge or discharge power of the BESS at bus $i$ at time $t$.
For daily operation of BESSs, the following constraints should be satisfied:
\begin{align}
    E_i^{end} &= E_i^{0}, t\in \mathcal{T} \label{SOC-2}\\
    \overline{P}_{BESS,i}^{dis} &\leq P_{BESS,i}^t \leq \overline{P}_{BESS,i}^{cha}, t\in \mathcal{T} \label{SOC-3}\\
    \underline{E}_{i} &\leq E_i^t \leq \overline{E}_{i}, t\in \mathcal{T}, \label{SOC-4}
\end{align}
where $E_i^{end}$ and $E_i^{0}$ are the stored energy of BESS at bus $i$ at the end of the optimization horizon and the initial state, respectively.   $\overline{P}_{BESS,i}^{dis}$ and $\overline{P}_{BESS,i}^{cha}$ are the power ramping limits of the BESS at bus $i$. $\underline{E}_{i}$ and $\overline{E}_{i}$ are the minimum and maximum energy stored in the BESS at bus $i$ and the latter also refers to the installed capacity of the BESS.

\begin{figure}[t!] 
\includegraphics[width=\linewidth]{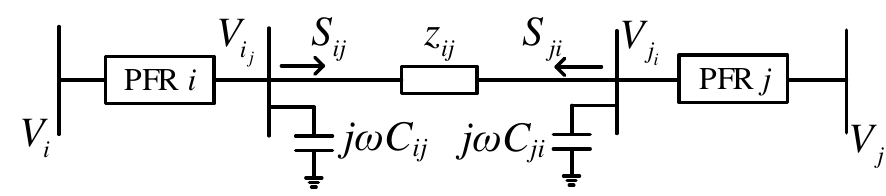}
\caption{Load flow model of PFR on a transmission line \cite{lin2017optimal}. }
\label{fig:PFR_branch}
\end{figure}

An example where PFRs $i$ and $j$ are installed at line $(i,j)$ is shown in Fig. \ref{fig:PFR_branch} \cite{lin2017optimal}. The branch terminal voltage of PFR $i$ at line $(i,j)$ is modelled by:
\begin{equation} \label{PFR-1}
    V_{i_j} = \gamma_{i_j}V_i,  \quad  \forall j \in \Omega_i,
\end{equation}
where $\gamma_{i_j} \in \mathbb{C}$ is the tuning variable of PFR $i$. 
The line flow is controlled by the terminal voltages $V_{i_j}$ and $V_{j_i}$, which makes the OPF problem with larger feasible region than the conventional one without PFRs. We focus on this feature to illustrate the grid flexibility brought by PFR rather than its hardware implementation. 

All the renewable sources are assumed to operate in the unity power factor mode which can be considered as negative loads. The renewable generation profile can be obtained based on the advanced forecasting techniques proposed in the literature \cite{6665108}.
We will investigate the influence of the renewable uncertainties as future work.

The steady-state power flow equations in the bus injection form can be described as follows:
\begin{align} 
    V_{i}^t\sum^N_{j=1}Y_{ij}^*(V_{j}^t)^* &= P_{G_i}^t+P_{wi}^t-P_{L_i}^t + P_{BESS,i}^t  \nonumber \\
    &+ {\rm{i}}(Q_{G_i}^t -Q_{L_i}^t) ,   \quad  \forall (i,j) \in \mathcal{L},  t\in \mathcal{T},\label{Powerflow}
\end{align}
where $P_{G_i}^t$ and $Q_{G_i}^t$ are the active power and reactive power output of the generator at bus $i$. Denote $P_{wi}^t$ as the active power output of renewable farm at bus $i$, $P_{L_i}^t$ and $Q_{L_i}^t$ as the active power and reactive power output of the load at bus $i$; and the superscript $^*$ means the conjugate of a complex number.

The voltage magnitudes, substation power supply, and line flow limits should be maintained within the predefined limits as follows:
\begin{align} 
\underline{V}_{i} &\leq |V_{i}^t| \leq \overline{V}_{i}, \quad \forall i \in \mathcal{N}, t\in \mathcal{T}  \label{voltage} \\
\underline{P}_{G_1} & \leq P_{G_1}^t  \leq \overline{P}_{G_1}, \quad t\in \mathcal{T} \label{substation-1}\\
\underline{Q}_{G_1} & \leq Q_{G_1}^t  \leq \overline{Q}_{G_1}, \quad t\in \mathcal{T} \label{substation-2} \\
\underline{S}_{ij} & \leq S_{ij}^t\leq \overline{S}_{ij}, \quad  \forall (i,j) \in \mathcal{L}, t\in \mathcal{T},    \label{lineflow}
\end{align}

In what follows, we assess the value of PFRs in reducing the total required BESS capacity while ensuring zero renewable power curtailment. 
%Let $\sum_{i=1}^{N_E}\overline{E}_{i} $ be the total BESS capacity installed in the system, where $N_E$ is the total number of BESSs. 
We first give a optimization benchmark where PFRs are not installed as follows:
\begin{align} 
&\text{minimize} \quad  \sum_{i=1}^{N}\overline{E}_{i}    \label{Benchmark} \\
&\text{subject to} \quad
\eqref{SOC-1} - \eqref{SOC-4}, \eqref{Powerflow} - \eqref{lineflow}, \nonumber
\end{align}
where the objective function is to minimize the total BESS capacity installed in the system. Note that this result gives the most optimal BESS capacity as we implicitly assume every node can install BESS. The setting excludes the influence caused by different BESS allocation schemes, so that we can focus on the influence of PFRs. 

Then we apply PFRs into the system and the optimization problem considering the operational limits of PFRs can be formulated as follows:
\begin{subequations} \label{MTOPF-BESS-PFR}
\begin{align} 
&\text{minimize} \quad  \sum_{i=1}^{N}\overline{E}_{i}    \label{M-1} \\
&\text{subject to} \nonumber\\ 
& \eqref{SOC-1} - \eqref{SOC-4}, \eqref{voltage} - \eqref{lineflow}, \nonumber \\
& V_{i_j}^t = \gamma_{i_j}V_i^t,  \quad \forall (i,j) \in \mathcal{L}, t\in \mathcal{T} \label{M-2} \\
&V_{i_j}^t\sum^n_{j=1}Y_{ij}^*(V_{j_i}^t)^* = P_{G_i}^t+P_{wi}^t-P_{L_i}^t+ P_{BESS,i}^t \nonumber  \\
&\qquad\quad\quad + {\rm{i}}(Q_{G_i}^t -Q_{L_i}^t) , \quad  \forall i \in \mathcal{N}, t\in \mathcal{T}\label{M-3}\\
&\underline{\beta}_{i_j} \leq \angle \gamma_{i_j}^t \leq \overline{\beta}_{i_j},\quad  \forall (i,j)\in \mathcal{L}, t\in \mathcal{T} \label{M-5} \\
&\underline{\gamma}_{i_j} \leq |\gamma_{i_j}^t| \leq \overline{\gamma}_{i_j},\quad  \forall (i,j)\in \mathcal{L},t\in \mathcal{T}, \label{M-6}
\end{align}
\end{subequations}
where \eqref{M-2} denotes the terminal voltages of PFR $i$ on line $(i,j)$. For the lines without PFRs,  $\gamma_{i_j}$ is set as 1. Eq. \eqref{M-3} represents the power flow equations.
%\eqref{M-5}-\eqref{M-6} specify the limits of dispatch power and reverse power flow through the transformer at substation, respectively. 
%\eqref{M-7} means that the branch power transfer must be restricted to the rating flow to satisfy the thermal limit. \eqref{M-8} refers to voltage constraints. 
Constraints \eqref{M-5}--\eqref{M-6} refer to the operational limits of PFRs. For a given time horizon, the optimal solutions to problem \eqref{MTOPF-BESS-PFR} and problem \eqref{Benchmark} give the minimum total BESS capacity required by the system to fully accommodate renewable energy when PFRs are installed or not.
%Note that this result gives the most optimal BESS capacity as we implicitly assume every node can install BESS. The setting excludes the influence caused by different BESS allocation schemes, so that we can focus on the influence of PFRs. 
The difference between the objective values of two optimization problems shows the contribution of PFRs in replacing BESSs. 
%Problem \eqref{Benchmark} and \eqref{MTOPF-BESS-PFR} are both non-convex due to the power flow equations, which makes the global optimal solution difficult to be found.

%The subtransmission system is connected with transmission network by the point of common coupling (PCC). Wind farms and BESSs are installed at the selected buses for power balance. The subtransmission network is characterized by high R/X ratio and therefore voltage overruns can be easily caused by this feature. 

\section{SDP Reformulation} \label{sec:methods}
In this section, we propose a methodology to address problems \eqref{Benchmark} and \eqref{MTOPF-BESS-PFR}. Since both of them are non-convex due to the power flow equations, convex relaxation is introduced to relax the problems into SDP forms, whose global optimality can be guaranteed. 

Recently convex relaxation techniques have been commonly used to convexify the OPF problem in power systems \cite{low2014convex1} and \cite{low2014convex2}.  In this paper, we adopt the SDP relaxation based on the bus injection model. A convexified problem is formulated according to the following steps.
Let $\mathbf{W^t} \triangleq \mathbf{V^t}\mathbf{V^t}^* \in \mathbb{C}^{2L \times 2L}, t\in \mathcal{T} $, be an auxiliary matrix. For each line $(i,j)\in \mathcal{L}$, let $\mathbf{V^t} \in \mathbb{C}^{2L}$ be the voltage vector composed of $V_{i_j}^t$ and $V_{j_i}^t$. We denote the diagonal entries of $\mathbf{W^t}$  as $W_{i_ji_j}^t = |V_{i_j}^t|^2,\forall i \in \mathcal{N},j\in \Omega_i$ and the off-diagonal entries as as $W_{i_jk_l}^t = V_{i_j}^t{V_{k_l}^t}^*, \forall i,j \in \mathcal{N},j\in\Omega_i,l\in\Omega_k$. By taking the magnitude of \eqref{PFR-1}, we define  $W_{i_ji_j}^t = \Gamma_{i_j}^t W_i^t$, where $W_i^t = |V_i^t|^2$, $\Gamma_{i_j}^t = |\gamma_{i_j}^t|^2, \forall i \in \mathcal{N}, j\in \Omega_i$.
%We introduce $\mathbf{W} \triangleq \mathbf{V}\mathbf{V}^* \in \mathbb{C}^{2E \times 2E}$ as the auxiliary matrix, where $\mathbf{V}  \in \mathbb{C}^{2E}$ is defined as the column voltage vector by stacking $V_{i_j}$ and $V_{j_i}$ for $(i,j) \in \mathcal{E}$. The diagonal entry of $\mathbf{W}$ is defined as $W_{i_j} = |V_{i_j}|^2,\forall i \in \mathcal{N},j\in \Omega_i$ and the off-diagonal entry is given as $W_{i_jk_l} \triangleq V_{i_j}V_{k_l}^*, \forall i,j \in \mathcal{N},j\in\Omega_i,l\in\Omega_k$. Denote $W_{i_j} = \Gamma_{i_j} W_i$ where $W_i = |V_i|^2$, $\Gamma_{i_j} = |1+\gamma_{i_j}|^2, \forall i \in \mathcal{N}, j\in \Omega_i$.
By defining $\mathbf{W^t}$ as the decision variable, problem \eqref{MTOPF-BESS-PFR} can be reformulated as follows:
\begin{subequations} \label{relaxed problem}
\begin{align}
&\text{minimize} \quad    \sum_{i=1}^{N}\overline{E}_{i}  \label{relax-1}  \\
& \text{subject to}   \nonumber \\
&\eqref{SOC-1} - \eqref{SOC-4},\eqref{voltage} - \eqref{lineflow}, \nonumber \\
& \sum^n_{j=1}Y_{ij}^*W_{i_jj_i} = P_{G_i}^t+{P}_{wi}^t + P_{BESS,i}^t , \nonumber\\
&\qquad\qquad -P_{L_i}^t+{\rm{i}}(Q_{G_i}^t-Q_{L_i}^t), \forall i \in \mathcal{N},t\in \mathcal{T} \label{relax-2}\\
& (\underline{V}_{i})^2 \leq W_i^t \leq (\overline{V}_{i})^2,  \quad \forall i \in \mathcal{N},t\in \mathcal{T} \label{relax-3}\\
& \mathbf{W^t} \succeq 0, \quad t\in \mathcal{T} \label{relax-4}\\
& \text{rank}(\mathbf{W^t}) = 1, \quad t\in \mathcal{T} \label{relax-5}\\
& \underline{\gamma}_{i_j}^2W_i^t \leq W_{i_ji_j}^t \leq \overline{\gamma}_{i_j}^2W_i^t, 
\quad \forall i \in \mathcal{N},k \in \Omega_{i}, t\in \mathcal{T}  \label{relax-6} \\
& \operatorname{Re}\{W_{i_ji_k}^t\}\tan{\underline{\theta}_{i_ji_k}} \leq \operatorname{Im}\{W_{i_ji_k}^t\}  \nonumber \\
&\leq \operatorname{Re}\{W_{i_ji_k}^t\}\tan{\overline{\theta}_{i_ji_k}}, \forall i \in \mathcal{N}, j\neq k \in \Omega_i, t\in \mathcal{T} \label{relax-7}\\
&\operatorname{Re}\{W_{i_ji_k}^t\} \geq W_i^t\underline{\gamma}_{i_j}\underline{\gamma}_{i_k}\cos{(\max{|\underline{\theta}_{i_ji_k}|,|\overline{\theta}_{i_ji_k}|})},\nonumber \\
&\qquad\qquad\qquad\qquad\qquad \forall i \in \mathcal{N}, j\neq k \in \Omega_i, t\in \mathcal{T}, \label{relax-8}
\end{align}
\end{subequations}
where \eqref{relax-2} refers to the power balance equations and \eqref{relax-3} indicates the voltage limits. Constraints \eqref{relax-6}--\eqref{relax-8} are the convexified reformulation of the original non-convex constraints \eqref{M-2} and \eqref{M-5}--\eqref{M-6}. 
%Therefore, the feasible regions of Problems \eqref{MTOPF-BESS-PFR} and \eqref{relaxed problem} are equal \textcolor{red}{(true?)}. 
Denote $\underline{\theta}_{i_ji_k}$ and $\overline{\theta}_{i_ji_k}$ as the limits of angle difference between two terminal voltages $V_{i_j}$ and $V_{i_k}$.
Constraint \eqref{relax-4} restricts the matrix $\mathbf{W^t}$ to be positive semi-definite at time $t$. Overall, problem \eqref{relaxed problem} becomes convex if the rank constraint \eqref{relax-5} is relaxed. 
SDP relaxation of OPF is shown to be usually reliable for OPF \cite{low2014convex1,low2014convex2} and the global optimal solution can be recovered from $\mathbf{W}^t, t \in \mathcal{T}$. In case the rank of $\mathbf{W}^t$ is larger than one, which means the relaxation is inexact, a penalty term can be added to the objective function, which has negligible effects on the optimality. Moreover, the benchmark problem \eqref{Benchmark} can also be relaxed in the similar way.
%A tree decomposition method can be adopted to reduce the computational burden of SDP relaxation with large-scale power systems \cite{lin2017optimal}.

\section{Case Study} \label{sec:case study}
\begin{figure}[t!] 
\includegraphics[width=\linewidth]{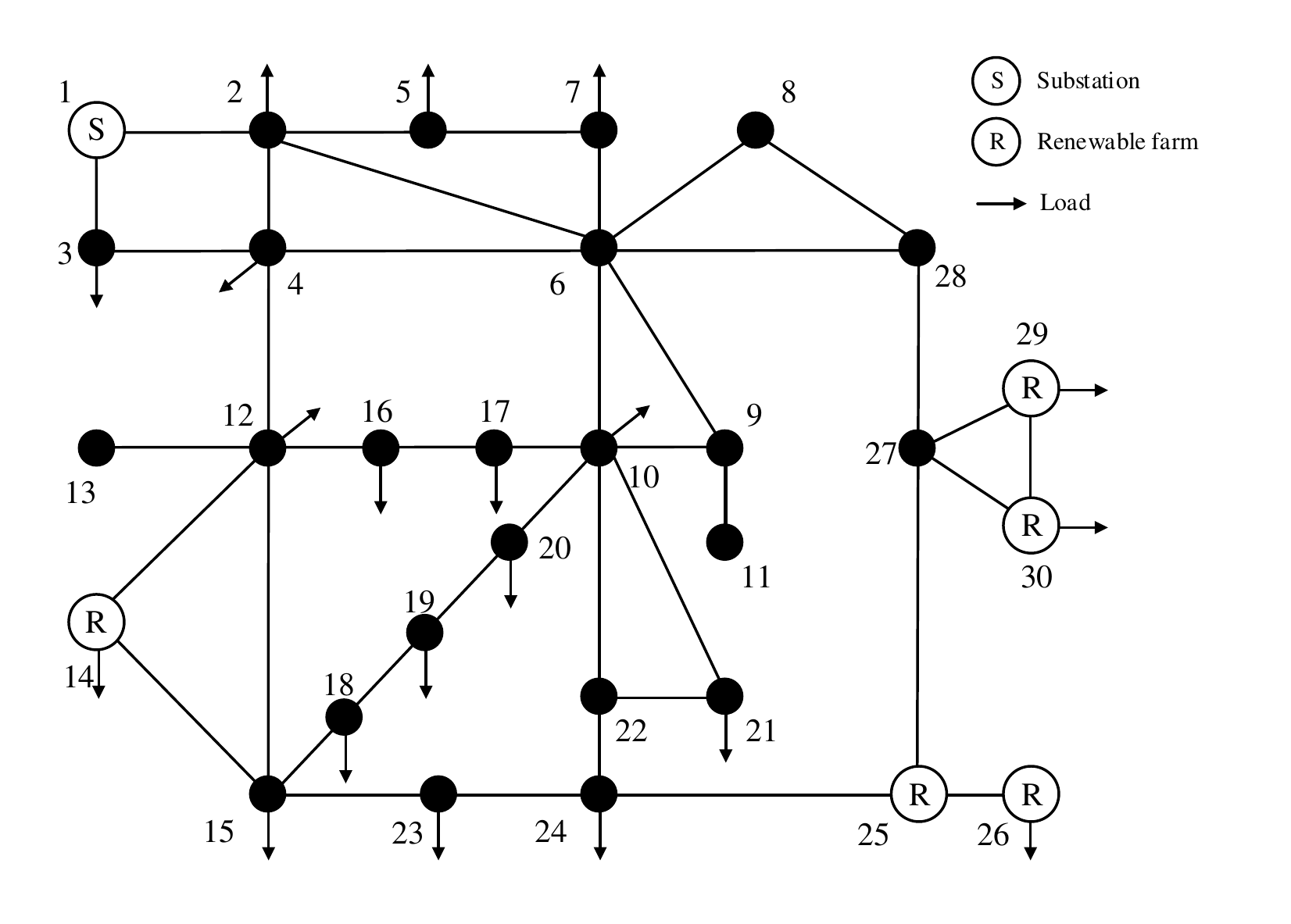}
\caption{Modified 30-bus test system. }
\label{fig:30busdiagram}
\end{figure}

\begin{figure}[t!] 
\includegraphics[width=\linewidth]{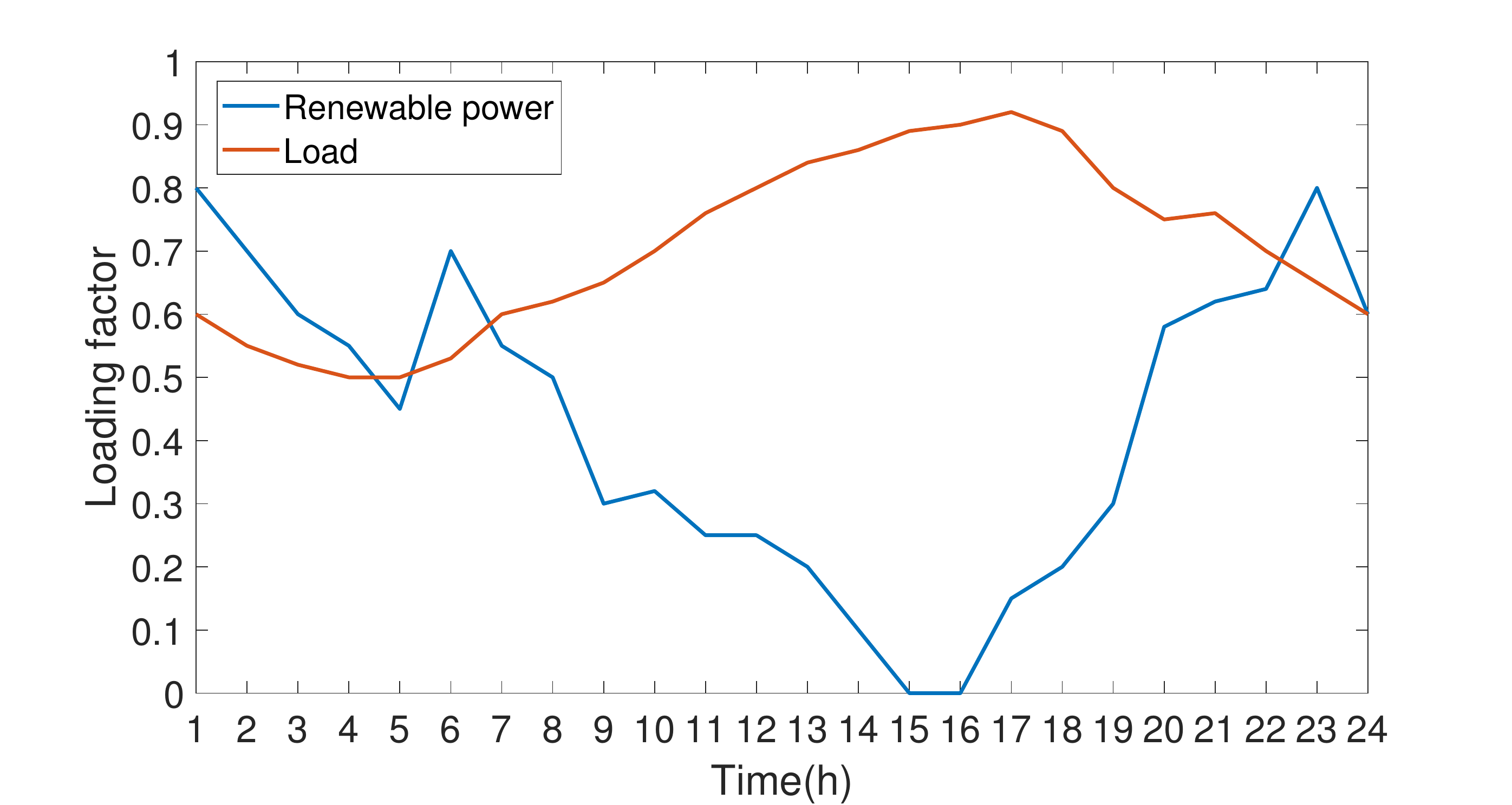}
\caption{24-h renewable generation and load profile }
\label{fig:resloadprofile}
\end{figure}
In the case study, we demonstrate the performance of the PFRs in reducing the required BESS capacity for zero renewable power curtailment.
The proposed multi-period optimization model is tested on a  modified IEEE 30-bus subtransmission system shown in Fig. \ref{fig:30busdiagram}. Basically we follow \cite{zimmerman2010matpower} to set up the network. The renewable farms are located at buses $\{14,25,26,29,30 \}$. We assume the capacities of the renewable farms are $20$ MW.
The BESSs are assumed to be installed as each bus since we focus on the performance of PFRs rather than the effects of BESS placement in this paper. In this way, the objective value gives the most optimal BESS capacity which can be regarded as the theoretical optimality.
A 24-hour time horizon is considered for the optimization model. The base case loading factors of load and renewable power at each hour are shown in Fig. \ref{fig:resloadprofile}, where the loading factors refer to the time-varying proportion of the base case value \cite{zimmerman2010matpower} for loads and the time-varying proportion of installed capacity for renewables, respectively.    %\textcolor{red}{which mean the proportion of the original base case load and the installed renewable capacity (not clear)}.  
For simplicity, all the renewable farm outputs have $100\%$ correlation following the loading factors given in  Fig. \ref{fig:resloadprofile}.  The program is implemented by CVX Toolbox \cite{grant2014cvx} and MOSEK \cite{mosek} in MATLAB.

\subsection{Case 1}

\begin{figure}[t!] 
\includegraphics[width=\linewidth]{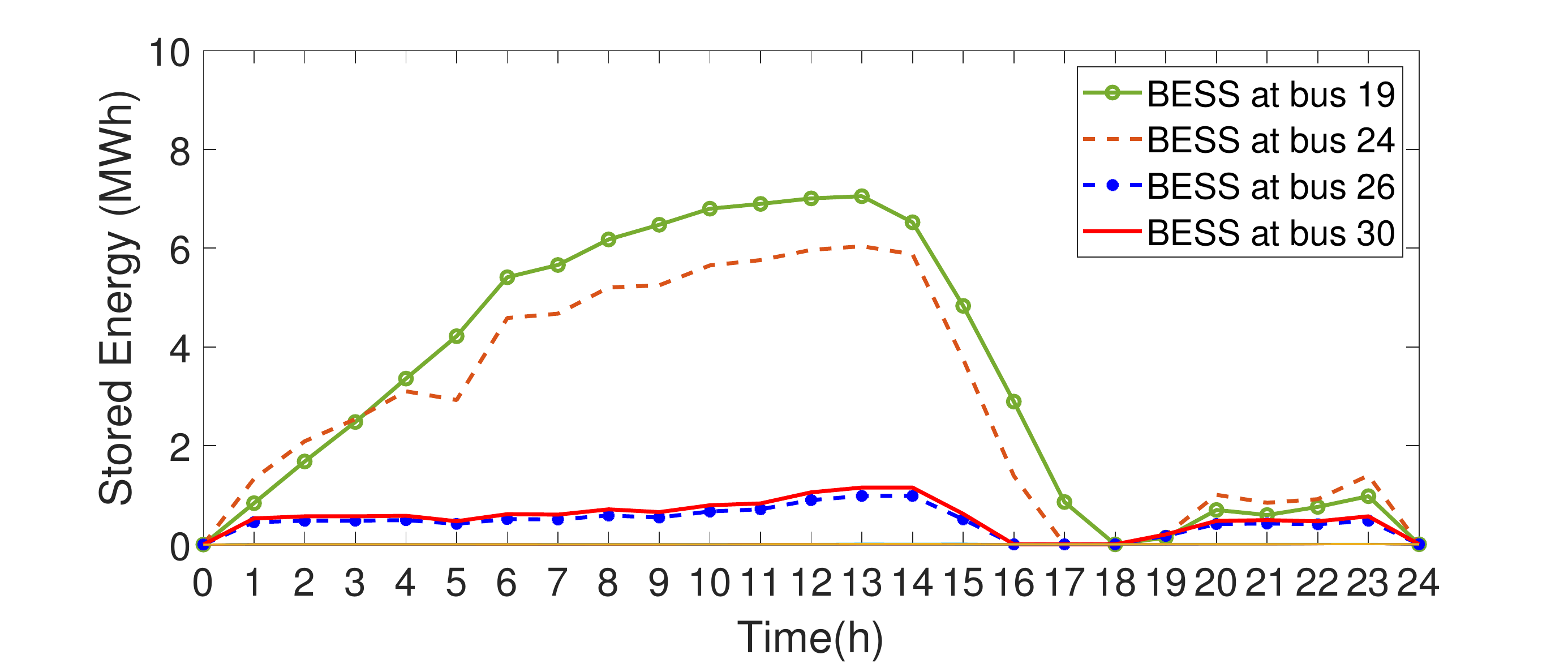}
\caption{Energy profiles of the BESSs without PFRs installed  }
\label{fig:socnoPFR}
\end{figure}

\begin{figure}[t!] 
\includegraphics[width=\linewidth]{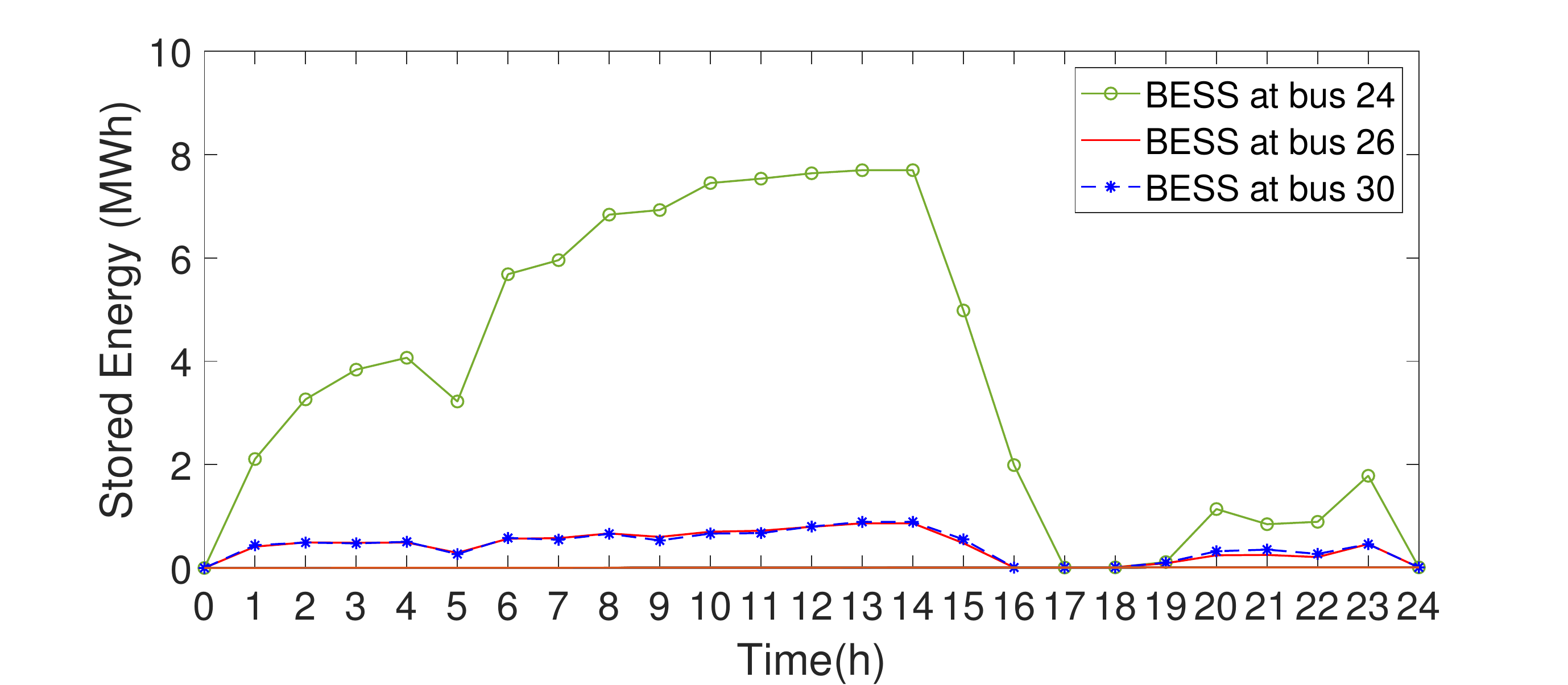}
\caption{Energy profiles of the BESSs with PFRs installed }
\label{fig:socwithPFR}
\end{figure}
First we consider the base case scenario without PFRs. In this case, only the BESSs provide active power support for voltage regulation within the predefined range. After solving the relaxation form of problem \eqref{Benchmark}, an optimal solution is derived as the minimum total required BESS capacity installed at all buses. The energy profiles contained in the BESSs are shown in Fig. \ref{fig:socnoPFR}. The peak values of the stored energy are equal to the minimum required BESS capacities installed at different buses. Therefore, the minimum values of the BESS capacity are $7.05$ MWh, $6.04$ MWh, $0.98$ MWh, $1.15$ MWh at buses $19, 24, 26$ and $30$, respectively. The required BESSs for other buses are close to zero and deemed negligible. The sum of the BESS capacity is $15.19$ MWh. It can be seen from the results that BESSs can effectively mitigate the voltage variations by charging power at the peak RES generation and discharging power at the peak load. The renewable power outputs are enlarged by $1.5$ times of the base case profile and the total capacity also increases to $18.53$ MWh. It is obvious that the cost of installing BESSs increases with the increasing penetration level of RESs. %\textcolor{red}{Other control devices or methods may be developed to realize the coordinated voltage control in subtransmission systems for reducing the cost of installing large size of BESSs. (what is the purpose of this sentence?)}

\subsection{Case 2}
To reduce the required capacity of BESSs for zero renewable power curtailment, we include a small number of PFRs for voltage control coordinated with BESSs. In this case, we first test the performance of PFRs with the basic renewable power and load profiles as above. 
%Since there are $41$ lines in the modified 30-bus subtransmission system, the optimization problem \eqref{relaxed problem} are solved  $41$ times \sout{with different Constraint \eqref{relax-3}}\textcolor{red}{(why?)}. 
%\textcolor{red}{We select the top five lines, where PFRs are installed, which lead to the minimum total values of BESS capacity (not clear)} 
According to the results by installing PFRs on a single line, the following five locations achieve the best performance: L24-25, L23-24, L27-29, L29-30, L25-26. We can see the corresponding values of the BESS capacity shown in Table \ref{table:1}. The minimum BESS capacity value is $12.85$ MWh when the PFRs are installed at line 24-25. The corresponding 24-hour energy profiles of the BESSs are shown in Fig.\ref{fig:socwithPFR}.
PFRs are able to reduce the BESS capacity by $15.40\%$ while satisfying the system limits in this case.
%, the PFRs on Lines 23--24 and 24--25 have the best performance, which indicates that the RES curtailment is normally caused by line flow congestion near the buses with RES farms. 
Similar to the base case, to further validate the performance of PFRs, the renewable power is enlarged by $1.5$ times compared to the base case profile and the corresponding minimum total capacities including all the BESSs are shown in Table \ref{table:2}. The percentage increase of the BESS capacity is quite small with respect to the value in the base case which implies the robustness of PFRs regarding the different penetration levels of renewables. 
%It is noted that for the extreme situation where all the lines with PFRs installed, BESSs are no longer needed. 
It is noted that for the extreme situation where much more lines with enough PFRs installed, BESSs are no longer needed. 
%Future work will study the coordinated planning model of PFRs and BESSs considering their costs.
For only one line with PFRs, the results show the remarkable effectiveness of PFRs and the robustness of PFRs has also been illustrated by adjusting the penetration level of renewables.

\begin{table}[!t]
\renewcommand{\arraystretch}{1.3} 
\centering
\caption{ Required BESS Capacity with PFRs Installed On a Single Line: Base Case (MWh)}
\label{table:1}
 \begin{tabular}{c| c| c| c| c| c }
\hline \hline
PFR location  & 23-24& 24-25 &25-26  & 27-29& 29-30    \\ 
  \hline
Required BESS Capacity  & 13.11 & 12.85 & 14.30 & 13.21 & 13.45  \\
\hline \hline
\end{tabular}
\end{table}

\begin{table}[!t]
\renewcommand{\arraystretch}{1.3} 
\centering
\caption{Required BESS Capacity with PFRs Installed On a Single Line: 1.5 times of RESs (MWh)}
\label{table:2}
 \begin{tabular}{c| c| c| c| c| c }
\hline \hline
PFR location   & 21-22& 23-24 &24-25 & 27-29& 29-30   \\
  \hline
Required BESS Capacity  & 14.87 & 13.65 & 13.15 & 14.11 & 13.77  \\
\hline \hline
\end{tabular}
\end{table}

% \subsection{Case 3}
% sensitivity analysis
% wind power penetration level vs BESS/PFR
% PFR numbers vs BESS

% Every line with PFRs \ Every bus with BESSs
% some of the lines with PFRs\ some of the buses with BESSs
% one PFR - "optimal placement"
% two PFRs - ... 
%-----------------------------------
\section{Conclusion} \label{sec:conclusion}
An optimization model that minimizes the required BESS capacity for accommodating the high penetration of renewables with the cooperation of PFRs is proposed in this paper. In the modern power systems, the injection of RESs and the peak load will cause massive voltage fluctuations which leads to extra demands of BESSs. Owing to the high cost of large-scale BESSs, PFRs are utilized in the proposed scheme to enhance the grid flexibility rather than having direct control of the node power injections. %The feasible regions of the branch power flows are enlarged by the PFRs, which can release power flow congestion.  
The performance of PFR is evaluated by the calculating difference between the minimum BESS capacity required for zero RES curtailment when PFRs are installed or not. 
The OPF-based problem is convexified by SDP relaxation and the operational constraints of PFRs and the BESSs are included. The simulation results show that PFRs are effective in reducing the BESS capacity without renewable power curtailment and $15\%$ of BESS capacity can be decreased when only one line installed with PFRs. The scenarios of PFRs installed at different lines are compared and the robustness of proposed method under high penetration of renewables is also studied. Future works include considering the uncertainties of RESs and loads, which may be achieved by developing a MPC-based coordinated control method for BESSs and PFRs.

\section*{Acknowledgment} \vspace{-0.1cm}
This research is supported in part by the National Natural Science Foundation of China, under Grant No. 51707170, and
the Theme-based Research Scheme of the Research Grants Council of Hong Kong, under Grant No. T23-701/14-N.
 \vspace{-0.2cm}
 
%\textcolor{red}{(for conference proceedings, use ``inproceedings'' instead of ``article'' in bibtex)}
\bibliographystyle{IEEEtran}
\bibliography{Reference}
\end{document}